\documentclass[prb,aps,twocolumn,showpacs]{revtex4}  
\usepackage{amsmath2000}
\usepackage{psfig}
\usepackage{graphicx}
\begin{document}
\def\CC{{\rm\kern.24em \vrule width.04em height1.46ex depth-.07ex
\kern-.30em C}}
\def\dwn{\downarrow}
\def\up{\uparrow}
\newcommand{\comm}[2]{\left[ #1, #2 \right]}
\renewcommand{\t}[1]{\tilde{#1}}
\def\qsqrt{{\sqrt{2} \kern-1.2em ^4}}
\def\up{\uparrow}
\def\dwn{\downarrow}
\def\L{{\cal L}}
\def\E{{\cal E}}
\def\F{{\cal F}}
\def\ga{{\cal G}[sl(2)]}
\def\H{{\cal H}}
\def\U{{\cal U}}
\def\P{{\cal P}}
\def\M{{\cal M}}
\def\W{{\cal W}}
\def\i{{\rm i}\,}
\def\d{{\rm d}\,}
\newcommand{\eps}{\varepsilon}
\newcommand{\beq}{\begin{equation}}
\newcommand{\beqa}{\begin{eqnarray}}
\newcommand{\eeq}{\end{equation}}
\newcommand{\eeqa}{\end{eqnarray}}
\newcommand{\nbeqa}{\begin{eqnarray*}}
\newcommand{\neeqa}{\end{eqnarray*}}
\newcommand{\bigfrac}[2]{\mbox {${\displaystyle \frac{ #1 }{ #2 }}$}}
\newcommand{\bra}[1]{\left\langle #1 \right |}
\newcommand{\ket}[1]{\left | #1 \right\rangle}
\newcommand{\fat}[1]{\mbox{\boldmath $ #1 $\unboldmath}}
\newcommand{\beqb}{\begin{eqblock}}
\newcommand{\eeqb}{\end{eqblock}} 
\newcommand{\ontop}[2]{\genfrac{}{}{0pt}{}{#1}{#2}}

\newenvironment{eqblock}[2]{\beq\label{#2}\begin{array}{#1}}{\end{array}
                                \eeq}
\newcommand{\dint}{\displaystyle \int\limits}
\newcommand{\tint}{\textstyle \int\limits}
\newcommand{\nbeqb}{\begin{neqblock}}
\newcommand{\neeqb}{\end{neqblock}} 
\newenvironment{neqblock}[1]{\[\begin{array}{#1}}{\end{array}\]}
\def\e{{\rm e}}
\def\id{{\rm 1\kern-.22em l}}
\newcommand{\dprod}{\displaystyle \prod\limits}
\newcommand{\dsum}{\displaystyle \sum\limits}
\newcommand{\bigchoose}[2]{\mbox {${\displaystyle{ #1 \choose #2}}$}}
\def\RR{{\rm
         \vrule width.04em height1.58ex depth-.0ex
         \kern-.04em R}}
\def\ZZ{{\sf Z\kern-.44em Z}}
\def\NN{{\rm I\kern-.20em N}}
\def\PACS{\par\leavevmode\hbox {\it PACS:\ }}

\title{Electrostatic analogy for integrable pairing force Hamiltonians}
\author{L. Amico$^{a}$, A. Di Lorenzo$^{a}$, A. Mastellone$^{a}$, A. Osterloh$^{a}$
, and R. Raimondi$^{b}$}
\affiliation{\hspace*{-2mm}${(a)}\,$ NEST--INFM $\&$ Dipartimento di Metodologie Fisiche e Chimiche (DMFCI), 
	Universit\`a di Catania, viale A. Doria 6, I-95125 Catania, Italy}

\affiliation{${(b)}\,$ NEST-INFM $\&$ Dipartimento di Fisica, 
Universit\`a di Roma Tre, Via della Vasca Navale 84, 00146 Roma, Italy.}

\begin{abstract}
For  the exactly solved reduced BCS model an 
electrostatic analogy exists; in particular it 
served to obtain the exact thermodynamic limit of the model 
from the Richardson Bethe ansatz equations. We present an electrostatic   
analogy for a wider class of integrable Hamiltonians with pairing 
force interactions. We apply it to obtain the exact thermodynamic 
limit of this class of models. 
To verify the analytical results, 
we compare them with numerical solutions of the Bethe ansatz equations for 
finite systems at half--filling for the ground state. 
\end{abstract}

\maketitle

\section{Introduction}

Pairing  force interactions have been 
successfully  employed  to explain phenomena in  different contexts 
such as superconductivity~\cite{TINKHAM}, 
nuclear physics~\cite{IACHELLO94},  
QCD~\cite{RISCHKE} and astrophysics\cite{ASTRO}. 
The original idea traces back to Bardeen, Cooper, and Schrieffer (BCS) 
who proposed  pairing   
of time-reversed electrons as the crucial mechanism  for superconductivity\cite{BCS}.
Physical implications of the corresponding  Hamiltonian have been extracted 
resorting to a variety of analytical or numerical 
techniques. The mean field ground state 
was shown to be exact  in the limit 
of large number of electrons where fluctuations can be 
neglected~\cite{TINKHAM,BCS,BOGOLIUBOV}. This constituted a great success of 
the BCS variational ansatz.   
However there are relevant physical situations  where approximations 
are not reliable and exact treatments are highly desirable. 
Fortunately, by properly choosing the pairing couplings, 
the model admits an exact solutions.\\
A simplified, but still non-trivial model is the ``reduced BCS model'' 
(BCS model, for brevity) which assumes a uniform pairing $g$ 
among  all the electrons within the   Debye shell.  
This model  is integrable~\cite{CAMBIAGGIO} and was diagonalized long 
ago by Richardson through Bethe Ansatz (BA)~\cite{RICHARDSON} 
(see also Ref.~\onlinecite{GAUDIN}).
Only very recently this  generated a lot 
of interest both in nuclear and condensed matter physics 
whose communities benefited from the 
simple algorithm of Richardson's BA solution to tackle the BCS model in the 
canonical ensemble~\cite{BRT,MASTELLONE,VONDELFT,NUCLEAR}. 
Much work has been done to merge the model in the schemes of
the Quantum Inverse Scattering (QIS) and Conformal Field Theory (CFT), 
which are modern arenas where quantum integrability and exact solutions 
can be treated on equal footing.     
These studies  allowed significant steps forward. 
QIS studies identified the Richardson 
BA solution  as  the {\it quasi classical limit} of the exact solution 
of (twisted) disordered six vertex 
models of the XXX-type~\cite{SKLYANIN-GAUDIN,AMICO}.  
This paved the way towards the exact evaluation  of correlation functions 
of the  BCS model~\cite{AO,AUSTRALIANS}. 
The field theoretical study of the  BCS model is due to  Sierra 
and coworkers. The first step 
was to relate the Richardson BA solution with WZNW-$su(2)_k$ 
models~\cite{SIERRA-CFT,SIERRA-REV}. This stimulated the discovery that 
the BA solution is the quasi classical limit 
of the Babujian's {\it off shell} BA of the (untwisted) disordered 
six-vertex model~\cite{OSBA,AMICO}. The field theoretical origin 
of the integrability of the BCS model has been 
clarified definitely~\cite{SIERRA-CS} to be 
a twisted Chern-Simons (CS) theory on a torus. 
The Richardson BA solution together with the 
underlying integrability of the theory arises from the 
Knizhnik-Zamolodchikov-Bernard equations (which are the Knizhnik-Zamolodchikov 
equations on the torus). 
The emergence of the CS theory is quite interesting and 
relates at a formal level the BCS theory of superconductivity 
with the Fractional Quantum Hall Effect (FQHE)~\cite{SIERRA-CFT} 
(also pointing out important differences).
In fact, the exact BCS wave function admits a ``Coulomb gas'' representation, 
which corresponds to the ``Coulomb gas'' representation of the 
Laughlin wave functions (plasma analogy)~\cite{PLASMA,SIERRA-CFT}.
Accordingly, also for the BCS model an electrostatic analogy does exist.
It  was illustrated first by 
Gaudin~\cite{GAUDIN-LARGE}: the Cooper pair energies are  
obtained as equilibrium positions of $N$ mobile charges in the background 
of $\Omega$ fixed charges and a uniform electric field of strength $1/g$.
This analogy was used by 
Richardson~\cite{RICH77} and Gaudin~\cite{GAUDIN-LARGE} 
to obtain the thermodynamic limit of the 
Richardson BA equations in two different approaches. 
Both reproduced the BCS mean-field gap equation, hence confirming the statement
of Bogoliubov~\cite{BOGOLIUBOV} that at $T=0$ the mean-field results 
become exact. 
Recently, Sierra and coworkers revived the attention on the 
thermodynamic limit of the BCS model, comparing 
Gaudin's long forgotten results with the numerical solution of the 
Richardson  BA equations for a finite number of  
particles~\cite{SIERRA-LARGE}.  

By generalizing the integrals of motion of the BCS model, 
the class of known integrable pairing-force Hamiltonians
could be enlarged considerably towards models with non-uniform 
interactions~\cite{ADO,RICHARDSON-NEW}
(for bosonic versions see the Ref.\onlinecite{DUKELSKY}). 
Here, the term ``non-uniform'' indicates that the interactions depend 
on the energy levels occupied by the interacting electrons.
As the uniform BCS model, also these models emerge from the quasi-classical
limit of the QIS approach for twisted disordered six vertex models of 
the XXZ-type\cite{AMICO,POGHOSSIAN}. The untwisted case, studied in 
the Ref.~\cite{HIKAMI}, corresponds to large effective pairing interaction.

In the present work we generalize the $2d$-electrostatic analogy to the 
class of integrable Hamiltonians obtained in Ref.~\onlinecite{ADO}. 
The thermodynamic limit of these Hamiltonians is obtained.  
We compare the obtained analytical results with numerical solutions 
of the BA equations.
The electrostatic analogy of the Richardson-Sherman equations
is a screening condition for a total electric field produced
by charges distributed in two-dimensional space. Accordingly, 
the electric field obeys a Riccati-type differential equations.

The paper is laid out as follows. In the next section we review 
the integrability of pairing force Hamiltonians that  arises 
from the underlying infinite dimensional Gaudin algebra $\ga$. 
In section III we present the electrostatic analogy,
review basic facts of $2d$-dimensional electrostatics
and apply them to obtain the thermodynamic limit of integrable 
Hamiltonians with non-uniform pairing couplings. 
At the end of the section, we obtain   Gaudin and 
Richardson's results  for the BCS model as a limiting case of our equations. 
Section IV is devoted to  conclusions.
In  Appendix A we sketch some mathematical aspects connected 
with the integrability 
of the models. In Appendix B  we present the connection with 
the Riccati equations by reviewing the work of 
Richardson~\cite{RICH77}. In Appendix C 
we collect details of the calculations. In Appendix D we discuss 
some features of the ground state and possible routes towards 
the study of the excitations.


\section{Exactly solvable  pairing models}\label{model}

The Hamiltonian for $N$ charged particles interacting through pairing force  
\begin{equation}
H = \sum_{i\sigma} \varepsilon_{i\sigma} n_{i\sigma} 
	+ \sum_{ij}
	U_{ij} \; n_{i} n_{j} - \sum_{ij} g_{ij} \; c^\dagger_{i\up} c^\dagger_{i\dwn} 
	c^{}_{j\dwn} c^{}_{j\up} ,
\label{general-model}
\end{equation}
has been proved~\cite{ADO} to be exactly solvable if the 
Coulomb   $U_{ij}$ and the pairing strength 
$g_{ij}$ are restricted to have the
following form
\begin{equation}
\left.
\begin{array}{rcl}
\label{couplings}
g_{ij} &=&  p \t{g} \bigfrac{\eps_{i}-\eps_{j}}{
	\sinh[p (u_{i}-u_{j})]} \;, \\
&& \\
4 U_{ij} &=& A - p \t{g} (\eps_{i} - \eps_{j}) \coth[p (u_{i}-u_{j})] \; , \\
\end{array}
\right\} \quad \mbox{for }i\neq j 
\end{equation}
where $\eps_i$ are the single-particle levels, while for 
convenience~\footnote{In  formulas~\eqref{couplings}  
instead of the single particle energies  the quantities $\eta_j= 
\varepsilon_j-g_{jj}/2+2 \sum_i U_{ij}$ enter. We fixed $g_{jj}$ and $U_{jj}$ 
such that $\eta_j=\eps_j$} we fix $g_{ii}=\phi$, $U_{ii}=A-\phi$, 
where \(2\phi=p\tilde{g} \sum_{i\neq j} (\eps_i-\eps_j)\coth{[p(u_i-u_j)]}\). 
The quantities $A$, $u$, and $u_i$  are arbitrary real parameters while 
$p$ can be real or pure imaginary.
To study the integrability of the Hamiltonian (\ref{general-model}) 
it is convenient to reformulate it  as a spin chain model. In fact it 
can be expressed  (up to a constant term) as 
\begin{eqnarray}
H &= \sum_i 2 \eps_i K_i^z  + 4 \sum_{i,j=1}^{\Omega} U_{ij} K_i^z K_j^z  
\nonumber \\
&\phantom{\sum_i\eps_i}-\frac{1}{2} \sum_{i,j=1}^{\Omega} 
 g_{ij} (K^+_i K^-_j + K_i^- K_j^+)   , 
\label{su(2)-model} 
\end{eqnarray}
where 
$
K_j^+ = c^\dagger_{j\up} c^\dagger_{j\dwn} , \quad K_j^- = 
\left(K_j^+\right)^\dagger \quad , 
K_j^z = \frac{1}{2} \left( n_{j\up} + n_{j\dwn} - 1 \right) ,
$ are  $su(2)$ operators.
The Hamiltonian (\ref{general-model}) acts non trivially only on 
the Hilbert space ${\cal H}$ of doubly occupied and empty level pairs; 
the singly occupied levels are 
excluded from the dynamics associated to (\ref{general-model}).
This phenomenon is commonly called the ``blocking of singly occupied levels''.
It reduces the problem of finding the spectrum of $H$
in the restricted Hilbert space ${\cal H}$ and it is an important ingredient for the 
integrability of the Hamiltonian (\ref{general-model}). 
Ultimately the latter resides 
in the algebraic connection of the pairing Hamiltonians with 
the infinite dimensional Gaudin algebra (see Appendix A)
$\ga=span \{ K^\pm(u), K^z(u) \, : \, u\in \CC\}$: 
the diagonalization of the Hamiltonian 
is  equivalent to the diagonalization of the correspondent of 
Casimir operator in  $\ga$, $K(u)$ . This is achieved either through algebraic 
or coordinate-wise BA, and corresponds to the 
quasi-classical limit of the inhomogeneous $XXZ$ 
model. The set of operators in involution are the residues 
of $K(u)$ and the  Hamiltonian is a  polynomial 
of these operators which are, by construction,  
the integrals of  motion of the model.
\\
In the following we will see these statements at work for the 
model (\ref{general-model}), (\ref{couplings}). 
Then we will recover the BCS model as a  limit of the pairing couplings.

\vspace*{8mm}
\begin{figure}[h]
\includegraphics[width=8cm,height=6cm]{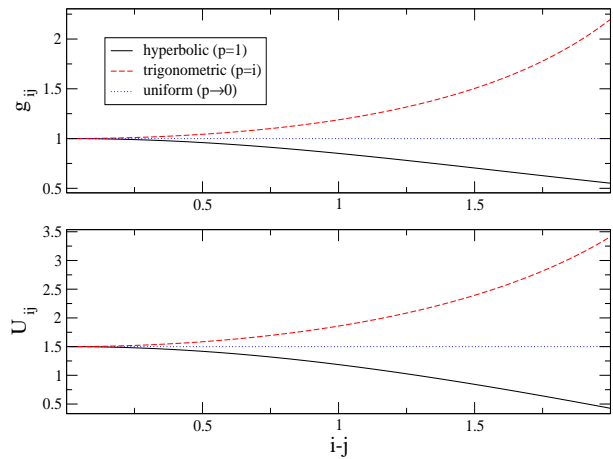}
\caption{Plot of the pairing  and Coulomb couplings in Eqs.(\ref{couplings})
with $\eps_i=i$, $u_i=\eps_i$, $A=2.5$ and $\tilde{g}=1$. In the numerics
done in this work these couplings have been considered for various values
of the rescaled $\tilde{g}$.}
\label{fig:couplings}
\end{figure}

\subsubsection{Non uniform couplings.}
We review the integrability of the pairing models~(\ref{general-model}).
The key observation is that the 
integrals of motion $\t{\tau}_i$ of the  model 
are residues of 
$\t{\tau}(u)$, defined in Appendix A, 
in $u=u_i$. 
In fact, $H$ can be written as 
\begin{eqnarray}
\label{eq:buildham}
 \hspace*{-10mm} H &=& \sum_i^\Omega 2\eps_i \; \t{\tau}_i 
+ A \sum_{i,j}^{\Omega} \t{\tau}_i\; \t{\tau}_j 
- \sum_i \phi_{i} {\bf K}_i^2 , \\
\hspace*{10mm}&&\t{\tau}_i={K}^z_i +\t{\Xi}_i \quad , \\
 \hspace*{-10mm} \t{\Xi}_i&:=& - p\t{g}
\sum_{j\neq i} \frac{{K}^+_i {K}^-_j +{K}^-_i {K}^+_j}
{2\sinh{[p(u_i-u_j)]}} \\
&&\hspace*{20mm}
+ {K}^z_i {K}^z_j \coth{[p(u_i-u_j)]} \nonumber
\end{eqnarray}
The $\tilde{g}\longrightarrow\infty$--limit of this Hamiltonian
was found in Ref.~\onlinecite{HIKAMI}.
From Eq.(\ref{eq:buildham}) it is evident that
$[H, \t{\tau}_j]=0$. 
By these integrals of motion, the 
model is connected with the anisotropic Gaudin Hamiltonians 
$\t{\Xi}_l$.  
The property~\cite{ADO}  
$[\t{\tau}_j, \t{\tau}_l] = 0$ for $j, l =1, \dots, \Omega$
comes from $[\tau (u), \tau (v)]=0$~\cite{SKLYANIN-GAUDIN}.
The common eigenstates of $H$ and $\t{\tau}_i$ are
\begin{eqnarray}\label{eigenstate}
\ket{\Psi} &=& \prod_{\alpha=1}^{N} K^+(e_\alpha)
\ket{0} 
\end{eqnarray}
where  $\ket{0}$ is the Fock vacuum and 
$N$ is the number of pairs.
The eigenvalues of the Hamiltonian $H$ and of the constants of the motion are 
\begin{equation}
{\cal E} = 
- p \t{g} \sum_{j=1}^{\Omega}\sum_{\alpha=1}^N \eps_{j} \coth{[p ( e_\alpha - u_j )]}
+ A N^2 
\label{eq:eigencharge}
\end{equation}
\begin{equation}
\tau_i = -\frac{\t{g}}{2} \sum_{\alpha=1}^N p \coth{[p (e_\alpha - u_i )]} 
+ \frac{\t{g}}{4} \sum_{\genfrac{}{}{0pt}{}{j=1}{j\neq i}}^{\Omega} p \coth{[p ( u_j - u_i )]} 
- \frac{1}{2} . 
\label{eq:gen-eigen}
\end{equation}
The quantities $e_\alpha$ are solutions of the following set of equations
\begin{equation} 
\frac{2}{\t{g}} + 2{\mathop{{\sum}}_{\beta=1 \atop \beta\neq\alpha}^{N}}
p\coth{[p ( e_\beta - e_\alpha )]}
-  \sum_{j=1}^\Omega p\coth{[p(u_j - e_\alpha)]} = 0 \;. 
\label{eq:generalized-richardson}
\end{equation}
which are the BA equations of the model~\eqref{general-model}.

\subsubsection{ Uniform couplings: the BCS model}
We now review the exact solution together with the integrability  
of the pairing Hamiltonian with uniform 
coupling constants. In this case 
the Hamiltonian(\ref{general-model}) reads 
\begin{eqnarray}
H_{BCS}=\sum_{i=1}^\Omega 2 \varepsilon_i K_i^z - \frac{g}{2} 
\sum_{i,j=1}^\Omega \left(K_i^+ K_j^- + K_i^- K_j^+\right)
\end{eqnarray}
$H$ can be directly diagonalized through coordinate-wise~\cite{RICHARDSON}
or algebraic~\cite{BRAUN} BA. Nevertheless its 
diagonalization can be achieved along the   lines  depicted 
above.
The integrals of motion $\tau_i$ of the  BCS model 
are the residue of 
$\tau(u)$
in $u=2 \varepsilon_i$. 
By these integrals of motion, the 
model becomes connected with isotropic Gaudin Hamiltonians $\Xi_l$.  
In fact, $H$ can be written as 
\begin{eqnarray}
H_{BCS} &=& \sum_{i=1}^\Omega 2\varepsilon_i \; \tau_i 
+  g \sum_{i,j=1}^\Omega \tau_i\; \tau_j +const.
\label{bcs} \\
\tau_i &=&  
 K_i^z + \Xi_i \quad , \\
\Xi_i &:=& - g \sum_{j\neq i}\frac{\vec{K}_i\cdot\vec{K}_j}{\varepsilon_i-\varepsilon_j} 
\label{isointmot}
\end{eqnarray}
where the spin vectors are: 
${\fat K}_j:= (K^x_j, K^y_j,K^z_j)$; 
$K^\pm_j=K^x_j\pm i K^y_j$.  
The eigenstates of both the BCS  Hamiltonian~\cite{RICHARDSON,GAUDIN} 
and its constants of motion $\tau_j$\cite{SKLYANIN-GAUDIN} are 
\begin{equation}
\ket{\Psi}_{BCS}= 
\prod_{\alpha=1}^{N} K^+(e_\alpha)
\ket{0} 
\end{equation}
The eigenvalues of the Hamiltonian and of the integrals of the motion are 
respectively
\begin{eqnarray}
{\cal E}_{BCS}&=&2 \sum_{\alpha=1}^N e_\alpha \;, \label{eigenbcs} \\  
\tau_i &=& -\frac{g}{2} \sum_{\alpha=1}^N  \frac{1}{ e_\alpha - \varepsilon_i } 
+ \frac{{g}}{4} \sum_{\ontop{j=1}{j\neq i}}^{\Omega}  
\frac{1}{ \varepsilon_j - \varepsilon_i } 
- \frac{1}{2} . \label{tau-BCS-eigenvalue} 
\end{eqnarray}
where $e_\alpha$ are solutions of the
Richardson-Sherman (RS) equations
\begin{equation} 
\frac{1}{g} +\sum_{j=1}^\Omega \frac{1/2}{e_\alpha-\varepsilon_j} 
-{\sum_{\ontop{\beta=1}{\beta\neq \alpha}}^N} 
\frac{1}{e_\alpha - e_\beta} = 0 \;
\label{richardson}
\end{equation}
(the parameters $e_\alpha$ are one half of the parameters $E_\alpha$ 
originally defined by Richardson~\cite{RICHARDSON}). 
The eigenstates, eigenvalues and the RS equations can be 
obtained in the $p\to 0$ limit of the BA 
equations~(\ref{eq:generalized-richardson}) with the identification  
\(u_i=\eps_i \), \(\t{g}=g\). By the same limit the anisotropic 
(XXZ) Gaudin model reduces to its  isotropic version (XXX).

\section{Electrostatic analogy for exactly solvable pairing models}

The RS equations \eqref{richardson} are stating that the 
total force acting on the unit charges located in $e_\alpha$, due to 
the $2d$ electric field\footnote{We recall that in two dimensions 
the electric field generated by a 
point charge $q$ having complex coordinate $z_0$ can be represented 
by the holomorphic function $E=q/(z-z_0)$.}
generated by the $\Omega$ charges $q_{fixed}=-1/2$ 
fixed in $\eps_i$, by the remaining $N-1$ mobile unit charges 
$q_{mobile}=+ 1$ and by the external constant electric field of 
strength $-1/g$ is zero.
In other words, the solutions of the RS equations 
correspond to the (unstable) equilibrium configurations of $N$ charges in the complex 
plane under the influence of a given electric field. 
This electrostatic analogy was first pointed out  by Gaudin\cite{GAUDIN}. 
It  allows  the exact access to the thermodynamic 
limit of the BCS model\cite{GAUDIN-LARGE} by exploiting the
formalism of complex analysis. The analytic structure of the 
ansatz electric field is prescribed by the positions of the charges.

We will now present an electrostatic analogy for the generalized 
BCS models~(\ref{general-model}), (\ref{couplings}). 

In  the variables 
\begin{equation}
z_\alpha:=\frac{\tanh{(p\, e_\alpha)}}{p} \quad \; ,\; \quad 
x_i := \frac{\tanh(p\,u_i)}{p}\;,
\label{change-variable}
\end{equation}
the Eqs.~(\ref{eq:generalized-richardson}) are 
algebraic 
\begin{equation} 
-\frac{q_{+}^{(0)}}{z_\alpha-1/p} -\frac{q_{-}^{(0)}}{z_\alpha+1/p} 
-\sum_{i=1}^{\Omega} \frac{1/2}{z_\alpha-x_i} 
+ \sum_{\genfrac{}{}{0pt}{}{\beta=1}{\alpha\neq \beta}}^{N}
\frac{1}{z_\alpha - z_\beta}
= 0\; , 
\label{eq:generalized-richardson2nd}
\end{equation}
where \(2q_{\pm}^{(0)}=\mp 1/p\tilde{g} -  [\Omega - 2( N -1)]/2\).
We shall assume that the parameters $u_i$ are given by 
\(u_i=f(\eps_i)\)  where $f$ is a monotonic function.
For the case $p=\i$ note that due to the $\pi$-periodicity of the 
transformation~(\ref{change-variable}) the $u_i$ can be restricted  to lie in 
$(-\pi/2,\pi/2)$.
Hence, the Bethe equations
for the generalized BCS models \eqref{eq:generalized-richardson} 
can be recast in the same form as the original RS 
equations Eq.~(\ref{richardson}), except that the constant 
external electric field
of strength $-1/g$ in the homogeneous case is replaced 
by the presence of two isolated charges $-q_{\pm}^{(0)}$ 
fixed at points $\pm 1/p$.\footnote{The analogy is by no means unique. 
In fact,  a general Moebius transformation $\mu (z)=r e^{i\phi} (z-\omega_1)/
(z-\omega_2)\; \omega_1,\omega_2 \in \CC$ preserves 
the algebraic structure of the Bethe equations~(\ref{eq:generalized-richardson2nd}), varying the locations of the charges.}
Note  that  for $p=\i$ the charges are complex. The  interpretation of this 
is discussed later on. 

\subsection{Basics of $2d$ electrostatics}
We sketch briefly the main ingredients of two-dimensional
electrostatics in terms of holomorphic functions in $\CC$ 
(i.e. real harmonic functions in $\RR^2$).

Let the electric field \({\mathbf E}=(E_x,E_y)\) be  associated to the 
complex number $E=E_x - \i E_y$. 
The Maxwell equations can then be summarized as
\beq
\partial_{\bar{z}} E = \frac{1}{2}
({\rm div}\vec{E} + \i ({\rm curl} \vec{E})_z)
= \pi( \rho + \i (\vec{j}_{mag})_z)  
\eeq
with real charge density $\rho$ and  real magnetic monopole current density 
$\vec{j}_{mag}$ in $z$-direction;  
the derivatives are
$\partial_z=(\partial_x -\i \partial_y)/2, 
 \partial_{\bar{z}}=(\partial_x +\i \partial_y)/2$.
Any integral over a closed curve $\Gamma$ gives
\nbeqa
\frac{1}{2\pi\i}\int_\Gamma \!\!\!d z\, E(z) &=& 
\frac{1}{2\pi}\int_\Gamma 
(E_x d y - E_y d x) -\i (E_x d x + E_y d y) \\
&\hat{=}& \frac{1}{2\pi} \int_\Gamma {\mathbf{E \cdot ds}} 
-\frac{\i}{2\pi} \int_\Gamma {\mathbf{E\cdot dl}} \\
&=& Q_\Gamma - \i I_\Gamma\;,
\neeqa
where  $Q_\Gamma$ and $I_\Gamma $ are the total electric charge 
and magnetic monopole current enclosed by $\Gamma$ (in the following 
we refer to $Q_\Gamma-\i I_\Gamma $ as charge).
The surface element $\mathbf{ds}$ is a vector perpendicularly
pointing outwards of $\Gamma$ with the same length as
the line element $\mathbf{dl}$. 

A charge $q$ contributes to the electric field with 
a simple pole with residue $q$. 
A line of charges gives a holomorphic
function on a Riemann surface with a branch cut along the charge line. 
The discontinuity of the field in crossing the cut gives the charge density 
\( \frac{1}{2\pi\i}\left(E_-(z) - E_+(z)\right)\), 
where $E_-(z)$ ($E_+(z)$) are the limiting values of the field when 
$z$ tends to the cut from the right (left) with respect to the  orientation
of the curve. 

\subsection{Thermodynamic limit} 

The thermodynamic limit of the model~\eqref{general-model}
 can be obtained 
in the following way: 
we first divide Eq.(\ref{eq:generalized-richardson2nd}) by $\Omega$ 
and define the (positive) charge densities
\nbeqa
\rho(x_j)&:=&\frac{1/2}{\Omega (x_{j+1}-x_j)}\\
\sigma(z_\alpha)&:=&\frac{1}{\Omega |z_{\alpha+1}-z_\alpha|}
\neeqa
To obtain a sensible thermodynamic limit, 
we assume i) that the pairing strength 
scales as $\t{g}=G/\Omega$, with fixed $G$; ii) that the Debye shell 
defined by the end points $\eps_1$, $\eps_\Omega$ does not depend 
on $\Omega$; iii) that the number of pairs increases with $\Omega$ 
according to $N=\nu\Omega$, where $\nu$ is the filling. 
In the limit $\Omega \to \infty$, 
Equations \eqref{eq:generalized-richardson2nd} then become
\beq
-\frac{Q_{+}^{(0)}}{z-1/p} -\frac{Q_{-}^{(0)}}{z+1/p} 
-\int_L \!\!d x\, \frac{\rho(x)}{z - x} 
+ \int_\Gamma \!\!|d z'|\frac{\sigma(z')}{z - z'}
= 0\; . \label{REgen:conti}
\eeq
where the integrals are meant in the sense of the principal 
value  ${\cal P}$. After the transformations (\ref{change-variable}) 
the transformed Debye shell  $L$ is still a segment of the real axis having 
end points $a_0=\tanh{(p\,u_1)}/p$, $b_0=\tanh{(p\,u_\Omega)}/p$; 
the isolated charges are $-Q_{\pm}^{(0)}=-\tfrac{1}{2}(\nu-1/2\pm 1/(pG))$;
the density $\rho$ is determined  
by the single-particle energy density $\rho_\eps$ as 
\(\rho(x)=\rho_\eps(\eps(x))/[f'(\eps(x))(1-p^2 x^2)]\), where 
\(\eps(x)=f^{-1}(\text{Arctanh}{(px)}/p)\), and it fulfills 
\begin{equation*}
\int_{\eps_1}^{\eps_\Omega} d \eps \rho_\eps (\eps)=
\int_L \!\!\!d x\, \rho(x) = 1/2\;.
\end{equation*}
The curve $\Gamma$ and the density $\sigma$ have to be determined, 
with the constraint 
\begin{equation*}
\int_\Gamma \!\!\! |d z| \sigma(z) = \nu \;.
\end{equation*}

The idea of Gaudin was to construct the total electric field $E(z)$ as a 
 function with an analytic structure prescribed by the actual 
charge distribution. He further assumed that eventual solutions 
of the  equation (\ref{REgen:conti}) are arranged 
in $K$ piece-wise differentiable arcs: 
\(\Gamma=\Gamma_\rho \cup \bigcup_{n=1}^{K} \Gamma_n\), where $\Gamma_\rho$ is 
the common support of $\rho(x)$ and $\sigma(x)$ which is not contained 
in the arcs \(\Gamma_{arc}\equiv \bigcup_{n=1}^{K} \Gamma_n\). 
For example, a charge distribution along the line $[a,b]$ 
in the complex plane with total charge $q$
and constant charge density   would lead to the total electric
field
\(
E(z)=({q}/|a-b|)\ln[(z-b)/(z-a)]
\)
which diverges at the end points of $[a,b]$.
Since in the case under consideration
 the electric field  has to
vanish at the end points of $\Gamma$, 
we need the charge density to vanish 
``sufficiently fast''  there.
An example of such a function could be any
\([(z-a)(z-b)]^\alpha \ln\frac{z-b}{z-a}\)
with any $\alpha\in \frac{1}{2}\NN$
or just \(\left[{(z-a)(z-b)}\right]^{\frac{\alpha}{2}}\) with odd $\alpha$. 
Both functions indeed vanish in $a$ and $b$ and have a branch cut 
along $[a,b]$.
The choice of the admissible functions is finally restricted by imposing 
that the free charges are distributed along a {\em finite} 
set $\Gamma$~\cite{GAUDIN-LARGE,SIERRA-LARGE}. In fact,
for $\alpha\not\in\frac{1}{2}\NN$ we either have infinite
branch cuts or divergences at the end points.
We want to stress that the run of the branch cut can be chosen
in an arbitrary way as long as it joins continuously the branch 
points $a$ and $b$. The run is fixed 
requiring   that it is an equipotential curve of $E(z)$.
The ansatz function for the total electric field is 
\beq\label{ansatz.gen}
\begin{split}
E(z)=&\,S(z)\left[
	\frac{Q_+}{z-1/p}+\frac{Q_-}{z+1/p}
+\int_L\!\!\!d x\, 
\frac{\varphi(x)}{z-x}\right], \\
S(z)=&\prod_{n=1}^K \sqrt{(z-a_n)(z-b_n)}.
\end{split}
\eeq
Note that the sign of the complex square root is uniquely given, 
once $\Gamma_{arc}$ is fixed, 
by $\sqrt{(z-a_n)(z-b_n)} = \sqrt{|z-a_n||z-b_n|} 
\exp{\frac{\i}{2} \arg_{\Gamma_n} (z-a_n) (z-b_n)}$.
The argument function $\arg_{\Gamma_n} (z-a_n) (z-b_n)$ is the sum 
of the angles between $z\in \CC$ and the two end points of $\Gamma_n$, 
which has a $2 \pi$--discontinuity crossing $\Gamma_n$. 
The unknown quantities are the end points $(a_n,b_n)$
of $\Gamma_{arc}$, $Q_\pm$ and $\varphi (x)$.
\\
To give a physical interpretation of such an ansatz function,
look at $S(z) Q_\pm$ and $S(z) \varphi(x)$ as
screened charges and charge density, respectively.
From the symmetry of the distribution of the fixed charges 
on $L$ and in $\pm 1/p$ it follows that
if $a_n$ has non-zero imaginary part, then $b_n=\bar{a}_n$, and 
we choose the orientation of $\Gamma_n$ from $a_n$ to $b_n$, 
with ${\rm{Im}}(a_n) < 0$. 
If $a_n$ is real, so must be $b_n$.
We argue however 
that this occurs only for distinct ``critical'' values
of the pairing coupling at which the imaginary part of complex 
conjugate $a_n$ and  $b_n$ vanishes and hence $a_n=b_n$.
The corresponding arc is then describing a closed curve.
Therefore we consider $S(z)$ to don't have cuts along the transformed
Debye shell $L$ (see Appendix \ref{app:excited}). We emphasize
that this does not mean that $\sigma(z)$ has no support on $L$.
A non-zero solution charge density on $L$ is accounted for by 
the screening density $S(x)\varphi(x)$.

After having determined the $a_n, b_n$, the curves $\Gamma_n$ are
found as equipotential curves of the 
total electric field Eq.(\ref{ansatz.gen}). 
The density $\sigma$ is then determined by the discontinuity of 
$E$ in crossing $\Gamma$. 
 
In the following, all the unknown quantities are obtained 
following a procedure {\it \`a la} Gaudin on which we report 
in the Appendix \ref{app:cauchy}.  
We will calculate the complex integrals involved in this procedure 
by exploiting basic knowledge of electrostatics.

Since the screened charges must tend to their bare values when 
the point $z$ is sufficiently close to the source, we have that 
\begin{subequations}\label{eq:screenedcharges}
\begin{align}
S(\pm 1/p)\;Q_\pm &= -\,Q_{\pm}^{(0)} \;,&& \label{eq:screenedcharges.a}
\end{align}
In a similar way, using $\sigma(z)-\rho(z)=(E_-(z)-E_+(z))/(2\pi \i)$
, we find
\begin{align}
S(x)\;\varphi(x) &= \sigma(x) - \rho(x) .  
\label{eq:screenedcharges.b}
\end{align}
\end{subequations}

Next, imposing that the electric field asymptotically goes as 
\(E(z)\sim 1/z^2\) (this is the leading term in the multi-pole expansion, 
since the total charge is zero), 
we obtain the following equations
\begin{align}
&\int_L dx\,x^n \varphi(x) = -\frac{1}{p^n} 
\left(Q_+ + (-1)^n Q_- \right)
, &&0\!\le\!n\!\!\le\!K\,.\hspace{15mm}\raisetag{6.5mm}\label{eq:endpoints}
\end{align}
We present a detailed derivation of these equations in Appendix \ref{app:cauchy}.
There we also show explicitly that the condition 
\(\int_\Gamma \sigma(z)|dz| = \nu\) is automatically fulfilled 
with Eqs.\eqref{eq:endpoints}.

Using Eqs.\eqref{eq:screenedcharges} we can express $Q_\pm$ and $\varphi(x)$ 
in terms of $S(\pm 1/p)$ and $S(x)$ respectively. 
The $\Gamma_n$ are finally obtained as equipotential curves
of the total electric field $E(z)$\cite{GAUDIN-LARGE,SIERRA-LARGE}
\beq
{\rm Re} \int_{a_n}^z \!\!\!d z'\, E_-(z') = \int E_x dx +E_y dy=0 \;,
\label{curve}
\eeq
The density $\sigma$ is determined by the discontinuity of $E$ 
along $\Gamma$, according to: 
\begin{equation}
\sigma(z) = \rho(z)+ \frac{1}{\pi}|E_-(z)|\;.
\end{equation}
It is worth noting that Eqs.\eqref{eq:endpoints} are 
a set of $K+1$ real equations for the $2K$ real parameters determining 
the end points of $\Gamma_n$. Gaudin stated that 
{\it seemingly} in any  case there is a finite number of 
solutions. From a physical perspective, though, we expect that the $K-1$ 
free parameters span a family of curves corresponding to a band of 
excitations of the system. 
This conjecture will be verified in forthcoming work\cite{FUTURE}. 

We summarize all the conditions found for the unknowns 
$a_n$, $b_n$, $\varphi$, $Q$
\begin{center}
\framebox{\makebox[8.5cm]{

\begin{minipage}{8.5cm}
\nbeqb{rcl}
S(z) &=& \dprod_{n=1}^K \sqrt{(z-a_n)(z-b_n)}\;,\\ \\
E(z)&=&S(z)\left[\bigfrac{Q_+}{z-1/p}+\bigfrac{Q_-}{z+1/p}+\dint_L\!\!\!d x\, 
\bigfrac{\varphi(x)}{z-x}\right]\;,\\ \\
\quad Q_\pm&=&-\bigfrac{\nu-1/2\pm 1/pG}{2S(\pm1/p)}
\\ \\
\varphi(x)&=&\bigfrac{\sigma(x)-\rho(x)}{S(x)} 
\\ \\
\quad\sigma(z)&=& \rho(z) +\bigfrac{1}{\pi}|E_-(z)|
\neeqb
The arcs are then determined by
$$\dint_L \!\! dx x^n \bigfrac{\rho(x)-\sigma(x)}{S(x)} = 
\bigfrac{Q_+ + (-)^n Q_-}{p^n} \,;
\ n =0,\dotsc,K$$
$$\Gamma_n 
: {\rm Re} \dint_{a_n}^z \!\!\!d z'\, E_-(z') = 0\;.$$
\end{minipage}
}}
\end{center}
As we stated above, the density $\rho$ of the transformed variables 
$x=\tanh{(pf(\eps))}/p$ is connected with the
given density $\rho_\eps$ of the single-particle levels 
via
\begin{equation*}
\begin{split}
\rho(x)&=\rho_\eps(\eps(x)) /[f'(\eps(x))(1-p^2 x^2)]\;,\\
\eps(x) &= f^{-1}(\text{Arctanh}(px)/p)\;.
\end{split}
\end{equation*}
The energy of the state is finally given by
\begin{align}
{\cal E} =&~ 
\alpha \nu^2 - G\int_L \!\!\!d x \rho(x)\eps(x) 
\int_\Gamma\!\!\!
|d z| \sigma(z) \frac{1-p^2 x z}{z-x}\,,\\
=&~\alpha \nu^2 + p^2 G \nu \int_{\eps_1}^{\eps_\Omega} \!\!\!d\eps\,\rho_\eps(\eps)\,\eps\,x(\eps)+ 
\nonumber \\
&~\text{ }- G\int_L \!\!\!d x\,\frac{\rho_\eps(\eps(x))\,\eps(x)}{f'(\eps(x))}
\int_\Gamma\!\!\!
|d z| \frac{\sigma(z)}{z-x}\,,
\end{align}
where $\alpha=A \Omega$.

\subsubsection{Comparison with the thermodynamic limit of the BCS model}
In the limit $p\to 0$, the two isolated charges $Q_\pm^{(0)}$ 
take an infinite value and 
they are displaced to infinity in such a way to give a uniform electric field
$1/G$. In this limit, the charges $Q_\pm$ behave as 
\begin{equation*}
\begin{split}
&Q_+ \sim \pm\left[-\frac{1}{2G}
-\frac{1}{2}
\left(\nu-\frac{1}{2} + \frac{1}{2G} \sum_{n=1}^K(a_n+b_n)\right) p\right] \
p^{K-1}\;, \\
&Q_- \sim \mp \left[\frac{1}{2G}
-\frac{1}{2}
\left(\nu-\frac{1}{2} + \frac{1}{2G} \sum_{n=1}^K(a_n+b_n)\right) p
\right] (-p)^{K-1}\;,
\end{split}
\end{equation*}
where we kept into account the dependence of the relative 
determination of $S(\pm 1/p)$ 
upon the number of arcs. 
Thus, the ansatz field reduces to 
\begin{equation}
E_{p\rightarrow 0}(z)\doteq E_0(z)=\,S(z)\int_L\!\!\!d x\, 
\frac{\varphi(x)}{z-x}\;, 
\end{equation}
Eq.\eqref{eq:screenedcharges.b} still hold, 
while Eqs.\eqref{eq:endpoints} simplify to: 
\begin{subequations}\label{eq:uniendpoints}
\begin{align}
&\int_L dx\,x^n \varphi(x) = 0
, \quad\quad 0\!\le\!n\!\!\le\!K-2\,, \\
&\int_L dx\,x^{K-1} \varphi(x) = -\frac{1}{G}\,, \\
&\int_L dx\,x^{K} \varphi(x) = -\nu+\frac{1}{2} -
\frac{1}{2G} \sum_{n=1}^K(a_n+b_n)\,,
\end{align}
\end{subequations}
The last two equations come from the fact that the field should go 
asymptotically as \(E_0(z)\sim -1/G+Q_T/z\) in this limit, 
since the total charge is now $Q_T = \nu - 1/2$. 
In the approach by Gaudin the last condition
was obtained by imposing \(\int |dz|\,\sigma(z) = \nu\).
This extracts the residue at infinity, which has to be such that the total
charge is zero. In our approach, we already started from a globally neutral 
system and hence this normalization condition for $\sigma(z)$ was
automatically fulfilled (see Appendix \ref{app:cauchy})

We compare our findings for integrable pairing models 
with the results for the BCS model obtained in 
Refs.~\onlinecite{GAUDIN-LARGE,RICH77} and reconsidered recently 
by Sierra and coworkers~\cite{SIERRA-LARGE}

\begin{center}
\begin{minipage}{8.5cm}
\nbeqb{rcl}
S(z) &=& \dprod_{n=1}^K \sqrt{(z-a_n)(z-b_n)}\;,\\ \\
E_0(z)&=&S(z)\dint_L\!\!\!d x\, 
\bigfrac{\varphi(x)}{z-x}\;,\\ \\
\varphi(x)&=& 
\bigfrac{\sigma(x)-\rho(x)}{S(x)}
\quad ; \ \sigma(z)= \rho(z) +\bigfrac{1}{\pi}|E_-(z)|
\neeqb
The arcs are then determined by 
\nbeqb{l}
\dint_L dx\,x^n \bigfrac{\sigma(x)-\rho(x)}{S(x)} = 0\, , 
\quad\quad n =0,\dotsc,K-2 \\ \\
\dint_L dx\,x^{K-1} \bigfrac{\rho(x)-\sigma(x)}{S(x)} = \bigfrac{1}{G}
\\ \\
\dint_L dx\,x^K \bigfrac{\rho(x)-\sigma(x)}{S(x)} 
-\bigfrac{1}{2G} 
\dsum_{n=1}^K (a_n+b_n)
= \nu-\bigfrac{1}{2}
\neeqb
$$\Gamma_n 
: {\rm Re} \dint_{a_n}^z \!\!\!d z'\, {E_0}_-(z') = 0\;.
$$
\end{minipage}
\end{center}

\subsubsection{Numerics and discussion of the solutions}
\begin{figure}[h]
\includegraphics[width=8cm,height= 6cm]{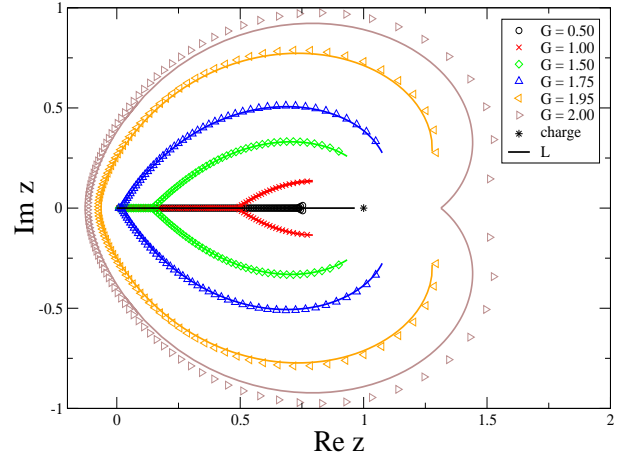}
\caption{
Plot of the pairing parameters $z$ in the complex plane for 
the hyperbolic model corresponding to $p=1$ in Eq.(\ref{couplings}).
The symbols refer to the numerical solutions of 
Eqs.\eqref{eq:generalized-richardson2nd}
 for $\Omega=200$, $N=100$, $u_i=\eps_i= 0,\dotsc ,2$ 
and uniform level spacing.  
Different colors 
and symbols represent different values of $G$, as shown in the
legend box. The lines are the plots of the curves determined 
by Eq.\eqref{curve}. If we choose the Debye shell to be $(-1,1)$ as
in the trigonometric case, the arc for $G=2$ closes at infinity.}
\label{fig:curve-hyper}
\end{figure}

To solve the BA equations ~(\ref{eq:generalized-richardson}), 
(\ref{eq:generalized-richardson2nd}) we have chosen the 
parameters $u_i=\eps_i$ at half filling $\Omega=2 N$. 
The corresponding interactions $U_{ij}$ and 
$g_{ij}$ are shown in Fig.\ref{fig:couplings}. 
In the trigonometric case the Debye shell ranges in 
$(-1,1)$;  in the hyperbolic case it ranges in $(0,2)$ 
for numerical convenience 
(see the caption of Fig.\ref{fig:curve-hyper}). 
The ground state of the system (which is the only state we consider in the numerics) 
is obtained by evolving  the $G=0$ state
\begin{equation}
\lim_{u \rightarrow 0} z_\alpha = x_\alpha \quad \alpha \in (1, \ldots, N).
\end{equation}
to a finite value of $G$. At certain value of $G$ the equations are singular since 
a couple of pairing parameters $z_{2\lambda-1}$ and $z_{2\lambda}$
coincide with  the energy level $x_{2\lambda-1}$. 
Such a  singular behaviour can be smoothed out using 
the   procedure developed by Richardson in 
Ref.~\onlinecite{RICH-NUM}. In the following we briefly summarize it.  
The corresponding divergences are removed by the following transformations   
\begin{equation}
z_{2\lambda-1} = A_{\lambda} -i B_{\lambda}, \quad
z_{2\lambda} = A_{\lambda} +i B_{\lambda},
\label{accoppio}
\end{equation}
and then
\begin{equation}
X_{\lambda}=A_{\lambda}-\frac{x_{2\lambda-1}+x_{2 \lambda}}{2}, \quad
Y_{\lambda} = - \bigfrac{B_{\lambda}^2}{\delta_{\lambda}^2 - X_{\lambda}^2}.
\label{outsing}
\end{equation}
where $\delta_{\lambda}=(x_{2\lambda}-x_{2 \lambda-1})/2$ 
is half the energy spacing between the corresponding energy values
(with whom they coincide if the pairing coupling strength $u$ is zero).

\vspace*{8mm}
By the first transformation Eqs.~(\ref{eq:generalized-richardson2nd}) 
can be written  in a form which is manifestly {\it real}, whereas the second 
transformation~(\ref{outsing}) removes the divergences. 

\begin{figure}[h]
\includegraphics[width=8cm,height= 6cm]{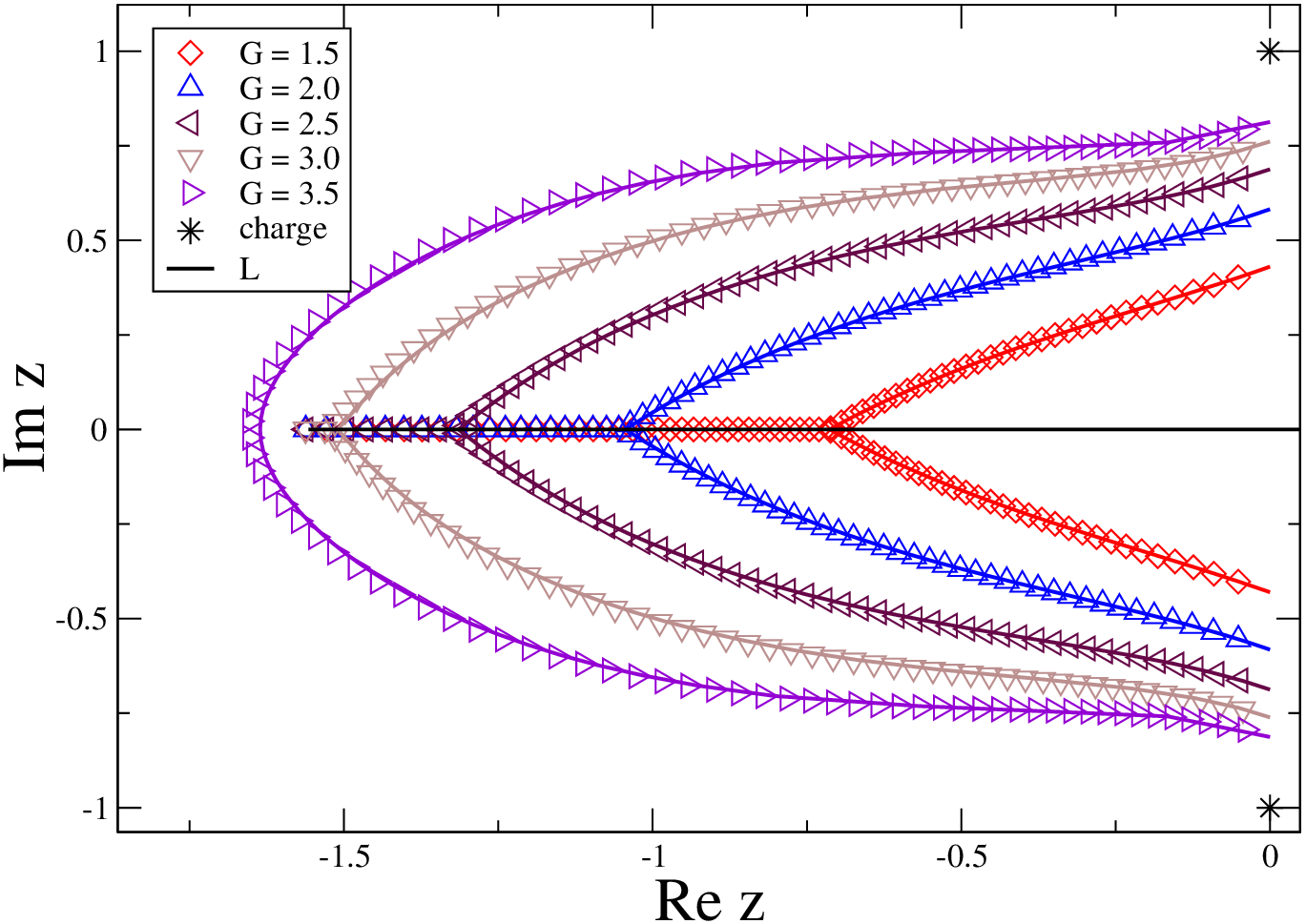}
\caption{
Plot of the pairing parameters $z$ in the complex plane for 
the trigonometric  model which correspond to $p=i$ in Eq.(\ref{couplings}).
The symbols refer to the numerical solutions of 
Eqs.\eqref{eq:generalized-richardson2nd}
 for $\Omega=200$, $N=100$, $u_i=\eps_i= -1,\dotsc ,1$ 
and uniform level spacing.  
Different colors 
and symbols correspond to different values 
of $G$, as shown in the legend box. 
The lines are the 
plots of the curves determined by Eq.\eqref{curve} 
.}
\label{fig:curve-trig}
\end{figure}

Now we discuss some features of the arcs shown in Figs.\ref{fig:curve-hyper} 
and \ref{fig:curve-trig}.
For $K=1$ (we recall that straight lines on the real axis are not 
counted by the index $n$), the end points are uniquely determined 
(if they exist) and the curve $\Gamma$ corresponds to the ground state\footnote{With {\em ground state} we mean that state connected 
with the ground state for $G=0$. Up to some ``critical'' value of $G$
this is the ground state of the system.}
of the system. 
In this case the end points of the arc, $\lambda \pm i \Delta$, are determined 
by the two coupled equations (obtained as linear combinations of 
Eqs.~(\ref{eq:endpoints}))
\begin{align*}
\int_L dx \frac{(1+px)\rho(x)}{\sqrt{(x-\lambda)^2+\Delta^2}} &= 
2Q_+=\frac{1/pG+\nu-1/2}{\sqrt{(\lambda-1/p)^2+\Delta^2}} \\
\int_L dx \frac{(1-px)\rho(x)}{\sqrt{(x-\lambda)^2+\Delta^2}} &= 
2Q_-=\frac{-1/pG+\nu-1/2}{\sqrt{(\lambda+1/p)^2+\Delta^2}}
\end{align*}
Why this set of equation corresponds to the ground state is discussed
in some detail in Appendix \ref{app:excited}.

In the trigonometric case ($p=\i$), the equations are both complex, 
and they are conjugate. 
Thus, in order to find $\lambda$ and $\Delta$ it suffices to solve 
separately real and imaginary part of one of them. 
In the hyperbolic case ($p=1$), assuming a uniform single-particle energy 
density $\rho_\eps(\eps(x))=\rho_0$ (i.e. $\rho(x)=\rho_0/(1-x^2)$) 
and a Debye shell spanning the interval $[0,2\omega_D]$, 
we have that the system above admits a real solution (and hence two 
coinciding end-points)
\begin{equation*}
\begin{split}
\Delta =& 0 \\
\lambda=&\sinh(2\omega_D)/\left(\cosh{2\omega_D}-\exp{\left[(\nu-1/2)/\rho_0\right]}\right)\;,
\end{split}
\end{equation*}
if \(1/(\rho_0 G) = 2 \omega_D\). This corresponds to the external 
curve in Fig.~\eqref{fig:curve-hyper}. 

\vspace*{8mm}
Situations corresponding to $K>1$ lead to  excited states.  
We point out that the  independent equations (\ref{eq:endpoints}) 
are in number of $K+1$ leaving undetermined $K-1$ parameters. 
This might be an evidence 
of ``bands'' of excitations above the ground state\cite{FUTURE}.  

\begin{figure}[h]
\includegraphics[width=8cm,height=6cm]{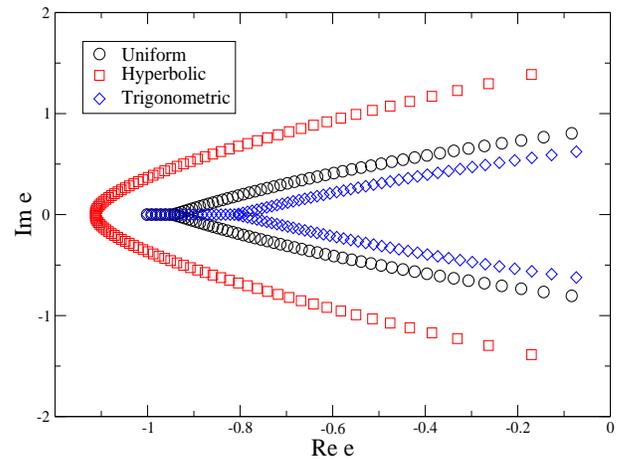}
\caption{
For an exemplary value of $G=2$ the solutions 
for $p=1,i$  corresponding to hyperbolic and trigonometric models respectively 
are compared with the solutions of the BA equation for the uniform case $p=0$ which 
are  the RS equations (the Debye shell ranges in $(-1,1)$ for all 
the three cases).}
\label{fig:comparison-curve}
\end{figure}
Looking at the behaviour of the quasi--momenta solutions of the 
BA equations, we can note that the more intense 
is the pairing constant $G$ the more evident 
is the tendency of the quasi--energies to be complex. In 
Fig.~\ref{fig:comparison-curve} we compare this tendency for the 
hyperbolic, trigonometric, and uniform BCS models. 
We found that the pairing tendency increases
from the trigonometric over the uniform to the hyperbolic model.
Looking at the explicit couplings plotted in Fig.~\ref{fig:couplings},
we should conclude that the pairing and Coulomb interaction in Eqs.~(\ref{couplings}) are competitive couplings.

\section{Summary and Conclusions}

We have discussed the electrostatic analogy  for the exactly solvable  
pairing force Hamiltonians found in Ref.~\onlinecite{ADO}. 
These models generalize 
the BCS model in that the coupling constants are not uniform 
(in the space of  quantum labels $i,j$, see Eqs.~(\ref{general-model})). 
The ordinary BCS model is 
obtainable as a  limit~\cite{ADO} $p\rightarrow 0$ of 
the models (\ref{general-model}), (\ref{couplings}). 
The electrostatic analogy  exists because  
the  BA equations for the ``rapidities'' $e_\alpha$
can be recast in a special algebraic form. For the class of models 
we deal with such an algebraic form is obtained  
by the change of  variables: 
$z_\alpha=\tanh(pe_\alpha)$, $x_i=\tanh(p\eps_i)$. After this transformation 
the Bethe equations express the condition for an equilibrium alignment of 
mobile charges $q_{mobile}=+1$ in the 
 $2d$--real or complex plane  in a neutralizing background 
of $q_{fixed}=-1/2$ fixed charges placed in the positions  $x_i$ and 
two further charges in the positions $\pm 1/p$.  
Remarkably, this analogy is 
very effective to obtain the exact thermodynamic limit 
(i.e. the large $N$ limit) of the models.  
In this limit the mobile charges (which are the solutions of the 
Bethe equations) arrange themselves along  equipotential curves 
$\Gamma=\bigcup_n^K \Gamma_n$ of the total 
electric field with vanishing charge density at its  extremities.
The electrostatic analogy presented in this work is by no means unique; 
the charge distribution can be transformed by the whole group of conformal 
Moebius transformations. 
We have obtained explicitly the thermodynamic limit 
for the ground state configuration at half--filling  
(see Fig.~(\ref{fig:curve-hyper}),~(\ref{fig:curve-trig})). We note, however, 
that the approach is also valid for excited states of the system, though
the explicit calculation can involve technical subtleties (discussed in 
Appendix \ref{app:excited}). \\ 
The  electrostatic analogy together 
with the constructive equations for the BCS model found by Gaudin 
and Richardson is demonstrated to  be obtained from the limit 
$p\rightarrow 0$ of the results presented here. A comparison 
between the hyperbolic, trigonometric, and uniform cases is seen in 
Fig.~(\ref{fig:comparison-curve}). Since this represent an inversion of 
the behaviour one would have expected from Fig.~(\ref{fig:couplings}) 
we conclude that  the pairing and Coulomb 
interactions in Eqs.~(\ref{couplings}) are  competitive couplings. 
The exact thermodynamics of  integrable pairing models is one of the 
future goals.

\acknowledgments 
We thank G. Sierra for helpful discussions. We further acknowledge 
constant support by G. Falci, R. Fazio, and G. Giaquinta.

\begin{appendix}
\section{The Gaudin algebra}
The Gaudin algebra 
${\cal G}[sl(2)]$ is constructed  from $sl(2)$. 
The  $sl(2)$  ``lowest''  weight module  is generated by the vacuum
vector $|0\rangle_j  $, 
$ 
K^-_j |0\rangle_j=0\;, \quad K^z_j |0\rangle_j=k_j |0\rangle_j
$
where $k_j$ is the ``lowest''  weight ($k_j=-1/2$ for spin $1/2$ which is 
the case considered for electrons). 
The infinite dimensional $\ga$ is generated by
\begin{equation}
K^\pm(\xi) := \sum_{j=1}^\Omega \phi(\xi-u_j) K^\pm_j
  \;, \quad 
K^z(\xi) := \sum_{j=1}^\Omega \psi(\xi-u_j)K^z_j
\; 
\label{gaudin-operators}
\end{equation}
where $\xi\in \CC$ and $u_j \in \RR$.
The module of ${\cal G}[sl(2)]$ is characterized by the vacuum 
$|0\rangle\equiv \otimes_{j=1}^\Omega |0\rangle_{j}$:
$
K^-(\xi) |0\rangle=0 
\; , \quad K^z(\xi) |0\rangle
= \kappa(\xi) |0\rangle \; ,
$
where 
$\kappa(\xi):=\sum_{j=1}^\Omega k_j \psi(\xi-u_j) $ 
is the lowest  weight  of ${\cal G}[sl(2)]$.
The element of $\ga$ corresponding to the $su(2)$ Casimir operator is \(
K(\xi):=K^z(\xi)K^z(\xi) + {\tfrac{1}{2}}\left( K^+(\xi)K^-(\xi)+K^-(\xi)K^+(\xi) \right)
\). The generating function of the integral of the motion of the 
pairing models is related to $K(\xi)$ through 
\begin{equation}
\tau (\xi) =-2\Lambda K(\xi)+K^z(\xi) + c(\xi) \id
\end{equation}
where $c(\xi)$ is a $\CC$-function and $\Lambda$ is a real parameter.
The  property  $[\tau(v),\tau(w)]=0$ arises from the quasi-classical limit 
of the $sl(2)$ QIS theory; this   is the ultimate reason for 
the integrability of the BCS model. 

In the present paper  
$\phi$ and $\psi$ are either rational or trigonometric/hyperbolic functions.

\subsection{Rational $\ga$}
In this case: $\phi(\xi)=\psi(\xi):=1/\xi$, 
$\kappa(\xi)\equiv k_0(\xi)=\sum_{j=1}^\Omega k_j/(\xi-u_j)$, and $\Lambda\equiv g$.
The operators~(\ref{gaudin-operators}) obey
\begin{eqnarray}
\nonumber
[ K^z(v),K^\pm(w)]&=&
\mp {\frac{K^\pm(v)-K^\pm(w)}{v-w}} \; ,\\ \nonumber
\quad 
[ K^-(v),K^+(w)] &=&2
{\frac{K^z(v)-K^z(w)}{v-w
}} \; ,
\label{gaudin-algebra}
\end{eqnarray}
where $v\neq w\;\in \CC$. 
The  Bethe equations are the Richardson-Sherman (RS) equations
\begin{equation}
\label{re}
k_0(e_\alpha) =\frac{1}{g } + \sum_{\ontop{\beta=1}{\beta\neq \alpha}}^{N} \frac{1}{ e_\beta- e_\alpha} 
\; , \quad \alpha = 1, \dots , N  \; .
\end{equation}
We note that RS equations~(\ref{re}) are intimately related to  
the algebraic structure of ${\cal G}[sl(2)]$ 
since they act as constraints on  the lowest  weight $k(E_\alpha)$. 
The difference between the BCS and Gaudin model 
results in a different constraint imposed on the lowest weight 
vector of ${\cal G}[sl(2)]$ which leads to different sets ${\cal E}$, 
${\cal E}'$ of solutions of the BA equations
($\E' $ is spanned by the solutions of ~(\ref{re}) 
when $g\to \infty$).
This fact has been used to extend the Sklyanin theorem for the Gaudin models
to the BCS model\cite{AO}.

\subsection{Hyperbolic/Trigonometric $\ga$}
In this case: $\phi(\xi):=p/\sinh[p\xi]$, 
$\psi(\xi):=p\coth[p\xi]$ , 
$\kappa(\xi)\equiv k(\xi)=\sum_{j=1}^\Omega p k_j \coth[p(\xi-u_j)]$and 
$\Lambda\equiv  \tilde{g}$.
The operators~(\ref{gaudin-operators}) obey
\nbeqb{l}
[K^z(v),K^\pm(w)]=
\mp p{\bigfrac{K^\pm(v)-\cosh[p(v-w)]K^\pm(w)}{\sinh[p(v-w)]}} \; ,\\ \\
{}[ K^-(v),K^+(w)] = 2p
{\bigfrac{K^z(v)-K^z(w)}{\sinh[p(v-w)]
}} \; ,
\neeqb
where $v\neq w\;\in \CC$. 

The BA equations for the corresponding Hamiltonian (\ref{general-model}) are
\begin{equation}
\label{re:ourmodel}
k(e_\alpha) =\frac{1}{\tilde{g}} +  \sum_{\ontop{\beta=1}{\beta\neq \alpha}}^{N} 
p\coth[p(e_\beta-e_\alpha)]\; , \; \alpha = 1, \dots , N  \; .
\end{equation}
Also in this case the BA equations act 
as constraints on  the lowest  weight $k(e_\alpha)$ of the trigonometric 
$\ga$. 

\section{The Richardson route and the Riccati equation}

In this appendix we first emphasize the key role played by 
the non linear differential equation of Riccati type in the electrostatic 
analogy of exactly solvable pairing models.\cite{RICCATI}
Then we sketch the procedure originally employed by 
Richardson~\cite{RICH77} to obtain the thermodynamic limit of the 
BCS model. 

In the non-uniform case the electric field is
\begin{equation}
{E}(z):=\frac{1}{2} {k}(z) + \frac{1}{2\tilde{g}} + 
\frac{p}{2} \sum_{\ontop{\beta=1}{\beta\neq \alpha}}^{N} 
\coth[p(e_\beta-z)]
\end{equation} 
with $k(z)$ as defined in the previous section.
In order to have the field obeying a Riccati-type equation,
it is crucial to map one of the isolated charges to infinity.
This is done by the transformation
\beq
q_i:=\exp{(2p \eps_i)}\,;\quad \zeta_\alpha:=\exp{(2p e_\alpha)}.
\eeq 
In the variables $q_i$ and $\zeta_\alpha$,  
the Riccati equation is
\begin{equation}
\frac{d{E}(z)}{dz}+ {E}^2(z) =\tilde{\tau}(z) + \tilde{f}(z)
\end{equation}
where $\tilde{\tau}_0(z):=1/\tilde{g}\sum_{j=1}^\Omega 
\tilde{\tau}_j/[q_j(q_j-z)]$ is a generating function of 
the eigenvalues of integrals of the motion 
of the model (\ref{general-model}) (see Eqs.~(\ref{eq:gen-eigen})) and  
\begin{eqnarray} 
\tilde{f}(z)=&&\frac{Q_0(Q_0+1)}{\tilde{g}^2} \\    
&&+\sum_{j=1}^\Omega \frac{1}{q_j -z} 
\left [\frac{k_j(k_j+1)}{q_j-z} -2 \frac{Q_0}{z}-2 \frac{P_0}{q_j} 
\right ] \nonumber
\label{riccati-trigo}
\end{eqnarray}
where $P_0:=2 Q_0 +\Omega -2 N +1/2$. 

For the isotropic limit $p\rightarrow 0$ the BCS rational case is recovered. 
The electrostatic field is here given by
(see RS Eqs.~(\ref{re}))
\begin{eqnarray}
E_0(z)&=&k_0(z) +\frac{1}{g } - 
\sum_{\ontop{\beta=1}{\beta\neq \alpha}}^{N} \frac{1}{ z-e_\beta} \\ 
&=&\frac{1}{g} +\sum_{j=1}^\Omega \frac{1/2}{z-\varepsilon_j} 
-{\sum_{\ontop{\beta=1}{\beta\neq \alpha}}^N} 
\frac{1}{z - e_\beta}  \; . \nonumber
\end{eqnarray} 
In this case the Riccati equation reads
\begin{equation}
\frac{dE_0(z)}{dz}+ E_0^2(z)=\tau(z) + f(z)
\end{equation}
where
\begin{equation}
f(z)=\frac{N^2}{g^2} +\sum_{j=1}^\Omega 
\frac{k_j(k_j+1)}{(z-\varepsilon_j)^2}
\label{riccati}
\end{equation}
where $\tau(z)$ is the generating function of the eigenvalues of 
integrals of the motion 
of the BCS model Eqs.~(\ref{tau-BCS-eigenvalue}).  
The Richardson BA equations are the zeros of the solutions 
of~(\ref{riccati}). The role of the Riccati equations in 
connection with the Gaudin models was also investigated in 
Ref.\onlinecite{RICCATI}. 

In both the rational or non-uniform cases it is evident that the 
Riccati equation plays an important role. 
This fact has been used by Richardson 
to derive the BCS theory from the expansion of the solution of the Riccati 
equation (in the rational case and $k_j=-1/2 \, \forall j$) 
in powers of $1/N$. In the following we 
briefly summarize the  Richardson's physical arguments.
The Richardson equations, via the electrostatic analogy,
can  be seen as a self-consistent evaluation of the field
of the fixed charges in the presence of the screening due
to the mobile charges. 
The strategy followed by Richardson is to attempt to eliminate
any reference to the mobile charges in the right hand side
of the above equation. In fact, using the Eqs.~(\ref{re}) the Riccati 
equation for the field  can be recast in the following form:
\begin{eqnarray}
\bigfrac{dE_0(z)}{dz}+E_0^2(z)&=&\sum_{i=1}^{\Omega}\frac{1/2}{(z-\epsilon_i)^2}+
\left[\sum_{i=1}^{\Omega}\frac{1/2}{z-\epsilon_i}+\frac{1}{g }\right]^2 
\nonumber \\ 
&&-2\sum_{i=1}^{\Omega}\frac{1/2}{z-\epsilon_i}
\sum_{\beta=1 }^{N} \frac{1}{\epsilon_i- e_\beta}.
\end{eqnarray}
In the last term on the right hand side there appears the field due
to the mobile charges at the location of the fixed charges. The
self-consistency condition, and the effective screening, enters here.
Richardson notices that the field of the mobile charges may be written
in the following way
\begin{equation}
\frac{1}{\epsilon_i- e_\beta}=\frac{1}{2\pi {\rm i}}\oint_{C}
{\rm d} z \frac{E_0(z)}{\epsilon_i -z}
\end{equation}
with the curve $C$ going around the singularities of $E(z)$ due to
the mobile charges. By using this, one gets a integro-differential
equation of the Riccati type
\begin{eqnarray}
\bigfrac{dE_0(z)}{dz}+E_0^2(z)&=&\sum_{i=1}^{\Omega}\frac{1/2}{(z-\epsilon_i)^2}+
\left[\sum_{i=1}^{\Omega}\frac{1/2}{z-\epsilon_i}+\frac{1}{g }\right]^2 
 \nonumber \\
+&\displaystyle \frac{1}{2\pi {\rm i}}&\oint_{C}
{\rm d} z' 
\sum_{i=1}^{\Omega}\frac{E_0(z')}{(z-\epsilon_i)(z'-\epsilon_i)}
 . 
\end{eqnarray}
The nice feature of the above equation is that any reference to the
mobile charges has disappeared. One may object that in the above equation
there is still a reference to the mobile charges via the curve $C$
entering the contour integral. In fact, by knowing the property
that the field has to satisfy at infinity one may observe that the integral
over a closed curve may be written as the sum of
two curves: the first is $C$ enclosing the singularities of the
mobile charges, while the second enclosing the singularities
of the fixed charges. Furthermore to evaluate the contour integral
along the curve extending at infinity, it is enough to know the first
term in the multi-pole expansion of the field. The final result is that
the integral over $C$ may be written in terms of the curve enclosing
the singularities of the fixed charges and the behavior of the field
at infinity only. The elimination of the mobile charges from the
effective equation for the field is closely in the spirit of the
Thomas-Fermi approach to the evaluation of the effective potential
due to the combined effect of fixed charges (nucleus in atoms, ions in metals)
and mobile charges (electrons). The integro-differential Riccati
equation plays in this context the role of the Poisson equation, which
becomes the Thomas-Fermi equation once the charge of the mobile
electron is written in terms of the potential.

The next trick used by Richardson is to expand 
the complex electric field $E_0(z)$ near the point at
infinity, in powers of $1/z$. This amounts to a multi-pole expansion.
The constant electric field gives the zeroth order term. The other terms
may be expanded upon using the geometric series. One easily gets
\begin{eqnarray}
\label{multipoles}
E^{(0)}_0 & =& -\frac{1}{g} \cr
E^{(1)}_0 &=& N-\frac{1}{2}\Omega \\ 
E^{(2)}_0 &=& \sum_{\beta =1}^N e_\beta-
\frac{1}{2}\sum_{i=1}^{\Omega}\epsilon_i \equiv \E-
\frac{1}{2}\sum_{i=1}^{\Omega}\epsilon_i \nonumber .
\end{eqnarray} 
The monopole and dipole terms give then the number of pairs and the energy.
To perform the thermodynamic limit, we rescale the coupling constant
as $g\rightarrow G/\Omega$ in order that the energy remains an extensive 
quantity in this limit (see section III B). The field is expanded as $E_0= E_{N} +E_1 + E_{1/N} +...$
where the subscript indicated terms of order $N$, $1$, and $1/N$, respectively.
We are assuming that $N$ and $\Omega$ keep a fixed ratio. 
 One may rewrite the Riccati equation order by order
\begin{eqnarray}
\label{riccatiexpansion}
E_{N}^2 & = & \left(\frac{\Omega}{G}+\sum_{i=1}^{\Omega}\frac{1/2}{z-\epsilon_i} \right)^2
-\sum_{i=1}^{\Omega}\frac{H_{N i}}{z-\epsilon_i} \\ 
2E_{N}E_1 &= &-\frac{d E_{N}}{d z}+\sum_{i=1}^{\Omega}\frac{1/2}{(z-\epsilon_i)^2}
-\sum_{i=1}^{\Omega}\frac{H_{1 i}}{z-\epsilon_i} \nonumber
\end{eqnarray}
where at each order in $N$
\begin{equation}
H_{i}=\frac{1}{2\pi {\rm i}}\oint_{C}
{\rm d} z \frac{E_0(z)}{\epsilon_i -z} .
\end{equation}
Following the analogous lines presented in section III, 
Richardson wrote down the solution
for $E_{N}$ as
\begin{equation}
E_{N}=-\frac{1}{2}Z(z)\sum_{i=1}^{\Omega}\frac{1}{Z(\epsilon_i )(z-\epsilon_i)}
\end{equation}
with $Z(z)=\sqrt{(z-a)(z-b)}$ having a branch cut between the points $a$ and
$b$. To see how the BCS limit is recovered, one has to expand
$E_{N}$ in  multiple's and insert the $E^{(m)}_{N}$ back 
into (\ref{multipoles}). A direct calculation shows that
\begin{eqnarray}
\label{multipoles2}
E^{(0)}_N & =& -\frac{1}{2} \sum_{i=1}^{\Omega} \frac{1}{Z( \epsilon_i )} \cr
E^{(1)}_N &=& -\frac{1}{2} \sum_{i=1}^{\Omega} \frac{1}{Z( \epsilon_i )} 
\left( \epsilon_i -\frac{a +b}{2}\right) \\
E^{(2)}_N &=& -\frac{1}{2} \sum_{i=1}^{\Omega} \frac{1}{Z( \epsilon_i )}
\left(\epsilon_i^2 -\epsilon_i \frac{a +b}{2} -\frac{1}{8}
(a-b)^2  \right) \nonumber.
\end{eqnarray} 
By setting $a=\lambda +{\rm i}\Delta$ and identifying the real and imaginary
parts of $a$ with the chemical potential and the energy gap, the first
two equations of (\ref{multipoles2}) reproduce the gap equation and the
condition for the chemical potential, while the third gives the ground state
energy. This completes the evaluation of the field in the leading order
 in $1/N$. To proceed further, one has to solve the second of
 (\ref{riccatiexpansion}), which is complicated by the way the
 field $E_1$ enters via the quantities $H_{1 i}$. Richardson showed, however,
 that there is an elegant way to circumvent the tackling of the integral
 equation.
An important point to notice is that the singularities of $E_{N}$
exhaust all the charges, so that the correction $E_1$ may have no poles
at the positions of the fixed charges. Inspection of the second of
Eq.(\ref{riccatiexpansion}) shows that $E_1$ has poles at the zeros of $E_{N}$.
However, $E_1$ cannot have poles because this would imply further charges.
To avoid this one has to impose that the right hand side of the second
equation of (\ref{riccatiexpansion}) has to vanish at the zeros of $E_{N}$.
There are $\Omega -1$ zeros of $E_{N}$ and one gets $\Omega -1$ linear equations
for the $\Omega$ unknown $H_{1i}$. One more equation may be obtained by
performing a multi-pole expansion in the Riccati equation. The Riccati
equation for $1/z$ terms reads
\begin{equation}
2E^{(0)}_0E^{(1)}_0=\frac{\Omega^2}{2 G}-\sum_{1=1}^{\Omega} H_{i}
\end{equation}
which may be rewritten as
\begin{equation}
N=\frac{G}{\Omega}\sum_{i=1}^{\Omega}H_i .
\end{equation}
By expanding in powers of $1/N$ one gets
\begin{eqnarray}
N &= &\frac{G}{\Omega}\sum_{i=1}^{\Omega}H_{N i} \cr
0&=  &\sum_{i=1}^{\Omega}H_{1 i} .
\end{eqnarray}
The second of the above equation provides the $\Omega$th linear equation
for the determination of the unknown $H_{1 i}$. These latter, once
determined, may be inserted back into the second of (\ref{riccatiexpansion})
to obtain the correction $E_1$.

\section{Electric field from contour integration}\label{app:cauchy}

In this appendix we present the derivation of 
Eqs.\eqref{eq:screenedcharges} and \eqref{eq:endpoints}
exploiting standard results of complex analysis.
Then we calculate the residue at infinity. Finally we discuss some
subtleties associated with the normalization condition for the
charge density $\sigma(z)$.

\bigskip
{\em 1. The contour integral $\dint \!\! dz E(z)/(z-z')$}

\begin{figure}
\includegraphics[width=8.5cm,height= 8cm]{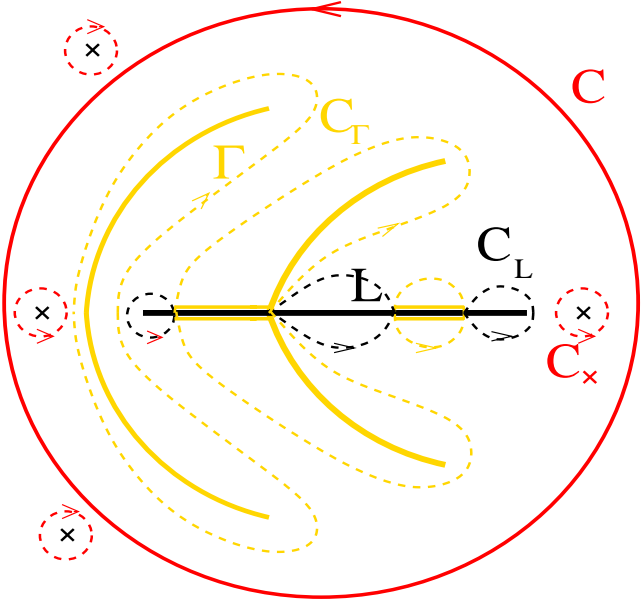}
\caption{The integration contour $L$ (full contour)
and its deformations are shown. The density 
$\sigma(z)$ of the ``solution charges'' is determined by 
evaluating once the inner part of $L$, and then the residue at infinity.
The contour integration $\int E(z)/(z-z')$ enclosing the arcs of 
$\Gamma$ extracts the charge density $\sigma(z)$ on these arcs: 
$\int_\Gamma |\d z| \sigma(z)/(z-z')$.}
\label{fig:contours}
\end{figure}

Our first aim is to write the last term in Eq.\eqref{REgen:conti} in 
terms of the ansatz field $E$ defined by Eq.\eqref{ansatz.gen}. 
To do this, we consider the contour $C$ shown in Fig.\eqref{fig:contours}. 
From the generalized Cauchy theorem, we know that the sum of the contour 
integrations of $E(z')/(z-z')$ inside $C$ is equal
to the negative residue of $E(z')$ at infinity (i.e. the charge at infinity),
which is zero in our case. 

The contours surrounding $\Gamma$ can be contracted to 
an integration along the right-hand side {\em minus} an integration along the
left-hand side of $\Gamma$. Since the discontinuity of $E(z)$ 
at $z\in\Gamma$ equals the charge density at $z$ times $2\pi\i$, this
gives
\beq\label{eq:justcharges}
\bigfrac{1}{2\pi\i}\int_{C_\Gamma} \d z \bigfrac{E(z')}{z-z'} = 
\int_{\Gamma} |\d z| \bigfrac{\sigma(z')-\rho(z')}{z-z'}\, .
\eeq
The integrals along the small half-circles 
at the end-points of $\Gamma$ tend to zero. If
we don't have an arc of $\Gamma$ on $L$ defined within 
the function $S(z)$, we obtain 
\begin{equation}\label{eq:cauchy}
\begin{split}
0=&\bigfrac{1}{2\pi\i}\oint_{C} \frac{E(z')}{z-z'}dz' = \frac{S(1/p) Q_+}{z-1/p}
+\frac{S(-1/p) Q_-}{z+1/p}  \\
&+\int_{\Gamma} |\d z| \bigfrac{\sigma(z')-\rho(z')}{z-z'}+\int_{L}\frac{S(x)\varphi(x)}{z-x}dx \stackrel{!}{=} 0\,.
\end{split}
\end{equation}
This finally leads to  
\begin{equation}
\begin{split}
\int_{\Gamma} \frac{\sigma(z')}{z-z'}|dz'|
&= - \frac{S(1/p) Q_+}{z-1/p} - 
\frac{S(-1/p) Q_-}{z+1/p} + \\
&-\int_{L\setminus\Gamma}\frac{S(x)\varphi(x)}{z-x}dx + 
\int_{L\cap\Gamma}\frac{\rho(x)}{z-x}dx \;.
\end{split}\label{eq:substitute}
\end{equation}
By substituting Eq.~(\ref{eq:substitute})
in the Bethe equations Eq.\eqref{REgen:conti}, we find the conditions in 
Eqs.\eqref{eq:screenedcharges}. 

Note that $\sigma(z)$ and $\rho(z)$ can have common support even though
the screening function $S(z)$ has no cut along $L$.
$L\cap\Gamma$ here means the common support of $\sigma(z)$ and $\rho(z)$.

\bigskip
{\em 2. Asymptotic behavior of $E(z)$ at $z=\infty$}

In order to find the remaining equations \eqref{eq:endpoints}, we expand 
the electric field in inverse powers of $z$, and impose that 
the leading order is $1/z^2$.  
\begin{equation*}
E(z)=z^{K-1} \sum_{j=0}^{\infty} c_j \left(\frac{1}{z}\right)^j 
\sum_{k=0}^{\infty} I_k \left(\frac{1}{z}\right)^k\;,
\end{equation*}
where $c_j$ are the coefficients of the expansion 
of \(\prod_{n=0}^K \sqrt{(1-a_n/z)(1-b_n/z)}\).
\begin{equation*}
I_j = Q_+ + (-1)^j Q_- + \int_L dx~x^j \varphi(x)\;.
\end{equation*} 
Equaling all powers $z^{i-1}$ for $i=0,\dotsc,K$ to zero, 
we obtain the triangular set of linear homogeneous equations for $I_j$: 
\begin{align*}
&c_0 I_0 = 0 \\
&c_0 I_1 + c_1 I_0 = 0 \\
&\vdots \\
&c_0 I_K + \dotsb + c_K I_0 = 0\;,
\end{align*}
which has the only solution (all $c_i\neq 0$) $I_0=\dotsb =I_K = 0$, 
which are the Eqs.\eqref{eq:endpoints}. 

\bigskip
{\em 3. The normalization of $\sigma(z)$ from Eqs.\eqref{eq:endpoints}}

We finally prove that the condition \(\int |dz|\,\sigma(z)=\nu\) 
is contained within Eqs.\eqref{eq:endpoints}. 

Contour integration of $E(z)$ yields
\nbeqa
\int |dz|\,\sigma(z) &=& \frac{1}{2\pi\i} \int_{C_\Gamma} \!\!\!dz\,E(z) + 
\int_{L\cap\Gamma} \!\!\!\!\!dz \rho(z)\\ \\
&=&-S(1/p) Q_+ - S(-1/p) Q_- \\
&& - \int_{L\setminus\Gamma} \!\!\!dx\,S(x) \varphi(x)+ 
\int_{L\cap\Gamma} \!\!\!\!\!dz \rho(z) \\ \\
&=&Q_+^{(0)}+Q_-^{(0)}+1/2 = \nu\,,
\neeqa
where we exploited the fact that \(\text{Res}\left\{E(z),\infty\right\}=0\),
which is the equation in \eqref{eq:endpoints} for $n=K$ if all the
equations for $0\leq n<K$ are fulfilled.

\section{Ground state and Excited states}\label{app:excited}

\begin{figure}
\includegraphics[width=8cm,height= 6cm]{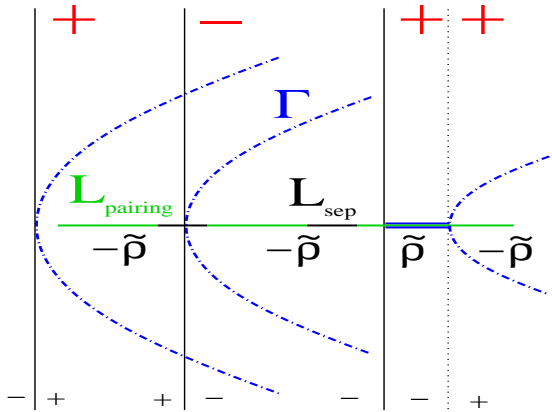}
\caption{Here it is indicated schematically how the signs of
the real square roots have to be chosen in order to yield
the sketched charge distribution. The function $S(z)$ consists of complex 
square roots and has a sign change at each crossing point of $\Gamma$
(dash-dotted grey curve) with the Debye shell (solid lines).
At the crossing point of the right-most part of $\Gamma$ a
sign change in the charge density ($\sigma=2 \rho$) is intended
and hence the sign change of $S$ located there (dotted vertical line) 
is ``shifted'' to represent the desired sign change $\rho\rightarrow -\rho$.
Grey parts of the Debye shell indicate a positive total charge density at $G=0$
and hence mobile charges; the black parts instead correspond to
a negative total charge density at $G=0$ and hence separated ``cells''.}
\label{fig:excited-states}
\end{figure}

We argue that one should not include arcs with end-points on $L$.
This is because the charges on such an ansatz arc have to be mobile.
The charges on the real axis however find themselves arrested 
in the ``cells'' flanked by two adjacent fixed charges. 
Therefore the only possibility for them to become mobile 
(in the thermodynamic limit) is to ``escape'' into the complex plane.
This however can happen only
for solution charges in neighbored ``cells''\cite{RICH-NUM}.
Zones of separated ``cells'' $L_{sep}$ are characterized by
$-\tilde{\rho}(x):=\sigma(x)-\rho(x)\leq 0$; $x\in L_{sep}$. We
can contribute for them, just taking $\tilde{\rho}$ as the density
of fixed charges. 
The complementary region $L_{pairing}:= L\setminus L_{sep}$ is thus
characterized by $\sigma(x)-\rho(x) > 0$ at $G=0$ and we let 
$\tilde{\rho}(x):=\rho(x)$.
In the simplest case we have $\tilde{\rho}(x)=\rho(x)$ everywhere
and also the modulus of the total charge is $\rho(x)$ in each point.
This means $\sigma(x)=0$ for $x\in L_{sep}$ and distinct connected
regions of neighbored ``cells'' with
$\sigma(x)=2\rho(x)$ for $x\in L_{pairing}$ at $G=0$ 
(see Fig.\ref{fig:excited-states}). This most simple situation
applies in particular to the ground state, 
which we have considered in the numerics.
The general situation can be attacked replacing $\rho(x)$ with 
$\tilde{\rho}(x)$ and the filling $\nu$ by 
\nbeqa
\tilde{\nu}&:=&\int_\Gamma |\d z| \tilde{\sigma}(z) \stackrel{G=0}{=}
		\int_{L_{pairing}}\hspace*{-6mm} \d x \tilde{\rho}(x) \\
&=& \nu- \int_{L_{sep}}\hspace*{-4mm} \d x \left[\tilde{\rho}(x) + \rho(x)\right].
\neeqa 
That the above argument applies also to the generalized class of BCS 
models discussed here, follows from the same structure of the 
equation after the transformation. 
But it can equally be seen from the electrostatic analogy,
since no two charges can penetrate each other due to the diverging
forces when sitting upon each other, which must be zero instead due to the
Richardson equations. These infinite forces can only 
be overcome if either at least one of the external charges diverges 
(i.e. the case G=0) or if two solution charges together approach a fixed 
charge (in the presence of a degeneracy $d$, more than two
solution charges are needed because the charge ratio 
$|q_{mobile}/q_{fixed}|=2/d$ is diminished by $d$).
We come back to this discussion after having presented the solution
of the problem.
Now we explain why the solution of Eq.(\ref{eq:generalized-richardson2nd}) 
corresponds to the
ground state of the system, which implies that up to the crossing 
point of $\Gamma$ with the Debye shell $L$ all ``cells'' between 
fixed charges are occupied. This implies that there we have
$\sigma = 2\rho$ and hence the total charge density is $-\rho$.
This sign change of the charge density below the crossing point
is already included in Eq.(\ref{eq:generalized-richardson2nd}), since we apparently replaced
\beq
\bigfrac{\rho(x)}{S(x)} \longrightarrow \bigfrac{\rho(x)}{\sqrt{(x-\lambda)^2+\Delta^2}}.
\eeq
Notice that the complex square root changes sign at the crossing point $x_c$:
indeed we have 
$$
S(x)=\left\{
\begin{array}{lcl}
\sqrt{(x-\lambda)^2+\Delta^2} & & x > x_c\\
&\mbox{for}&\\
-\sqrt{(x-\lambda)^2+\Delta^2} & & x < x_c
\end{array}\right. .
$$
This means that in Eq.(\ref{eq:generalized-richardson2nd}) the charge density is $-\rho(x)$ for
$x<x_c$ and $\rho(x)$ for $x > x_c$, which correspond to the ground state.
For general excited states, which might also include more than one arc,
we must furnish the real square roots with the proper signs as
indicated in Fig.\ref{fig:excited-states}.
As long as all the arcs cross the real axis in
the same interval of $L_{pairing}$ where they are generated from,
this procedure can be implemented in the formalism described in this
work. Otherwise the crossing points could be determined numerically.

\end{appendix}



\begin{thebibliography}{99}
%
\bibitem{TINKHAM}
M.~Tinkham.
\newblock {\em Introduction to Superconductivity}.
\newblock McGraw-Hill, New York (1996).
\newblock 2nd edition.
%
\bibitem{IACHELLO94}
F.~Iachello, Nucl. Phys. A {\bf 570}, 145 (1994).
%
\bibitem{RISCHKE}D.H. Rischke and R.D. Pisarski, 
{\em "Fifth Workshop on QCD"}", Villefranche, nucl-th/0004016.
%
\bibitem{ASTRO}
H.~Heiselberg and M.~Hjorth-Jensen, Phys. Rep. {\bf 328}, 237 (2000).
%
\bibitem{BCS}J. Bardeen, L.N. Cooper, and J.R. Schrieffer, Phys. Rev. 108,
	1175 (1957).
%
\bibitem{BOGOLIUBOV}
N.N. Bogoliubov, Nuovo Cimento {\bf 7}, 794 (1958).
%
\bibitem{CAMBIAGGIO}
M.C. Cambiaggio, A.M.F. Rivas, and M.~Saraceno, Nucl. Phys. A {\bf 624}, 157
  (1997).
%
\bibitem{RICHARDSON}
R.W.  R.W. Richardson, Phys. Lett. {\bf 3},  277 (1963); 
R.W. Richardson and N. Sherman, Nucl. Phys. {\bf 52}, 221 (1964); {\bf 52}, 253 (1964).
%
\bibitem{GAUDIN}
M.~Gaudin, J.  Physique {\bf 37}, 1087 (1976).
%
\bibitem{BRT} 
	C.T. Black, D.C. Ralph, and M. Tinkham, 
	Phys. Rev. Lett. {\bf 74}, 3241 (1995); {\bf 76}, 688 (1996);
	{\bf 78}, 4087 (1997). 
%
\bibitem{MASTELLONE}
	A.~Mastellone, G.~Falci, and R.~Fazio, Phys. Rev. Lett. {\bf 80}, 4542 (1998); 
	A. Di Lorenzo,  R. Fazio, F.W.J. Hekking, G. Falci, 
	A. Mastellone, and G. Giaquinta, Phys. Rev. Lett. {\bf 84}, 550 (2000);
	G. Falci, A. Fubini, and A. Mastellone, Phys. Rev. B {\bf 65}, 
	140507R (2002).
%
\bibitem{VONDELFT}
	J. von Delft and D.C. Ralph, Phys. Rep. {\bf 345}, 61 (2001). 
	M.~Schechter, Y. Imry, Y. Levinson, and J. von Delft, 
	Phys. Rev. B {\bf 63}, 214518 (2001).
%
\bibitem{NUCLEAR} J. G. Hirsch, {\it et al.} {\it  nucl-th/0109036}.
%
\bibitem{SKLYANIN-GAUDIN}
	E.K. Sklyanin, J. Sov. Math. {\bf 47}, 2473 (1989).
%
\bibitem{AMICO} L.~Amico, G.~Falci, and R.~Fazio, 
	J. Phys. A {\bf 34}, 6425 (2001).
%
\bibitem{AO} L. Amico and A. Osterloh, Phys. Rev. Lett. {\bf 88} 127003 (2002).
%
\bibitem{AUSTRALIANS} J. Links, H.-Q. Zhou, R.H. McKenzie, M.D. Gould 
	Phys. Rev. B {\bf 65}, 060502R (2002).	
%
\bibitem{SIERRA-CFT}
G.~Sierra, Nucl. Phys. B {\bf 572}, 517 (2000).
%
\bibitem{SIERRA-REV} G. Sierra, 
	{\it Proceedings of the NATO Advanced Research 
	Workshop on Statistical Field Theories, Como 2001}, 
	Eds.: A. Cappelli and G. Mussardo  
	(Academic Press, Cambridige 2001).
%
\bibitem{OSBA} H.M. Babujian, J. Phys. A {\bf 26}, 6981 (1993);
	H.M. Babujian and R. Flume, Mod. Phys. Lett.{\bf 9} 2029 (1994).
%
\bibitem{SIERRA-CS} M. Asorey, F. Falceto, and G. Sierra,
         Nucl.Phys. B {\bf 622} 593  (2002). 
%
\bibitem{PLASMA} R.B. Laughlin, Phys. Rev. Lett. {\bf 50}, 1395 (1983) 
%
\bibitem{GAUDIN-LARGE} M. Gaudin, 
	{\it Travaux de M. Gaudin. Mod\`eles Exactement r\'esolus}
	(Les Editions de Physique, France 1995).
%
\bibitem{RICH77} R.~W. Richardson, 
	J.~Math.~Phys. {\bf 18}, 1802 (1977).
%
\bibitem{SIERRA-LARGE} J.M. Roman, G. Sierra, and J. Dukelsky, 
	{\it cond-mat/0202070}.
%
\bibitem{ADO} L. Amico, A. Di Lorenzo, and A. Osterloh, 
	Phys. Rev. Lett. {\bf 86}, 5759 (2001); Nucl. Phys. B 
	{\bf 614}, 449 (2001).
%
\bibitem{RICHARDSON-NEW} R.W. Richardson, {\it cond-mat 0203512}.
%
\bibitem{DUKELSKY} J. Dukelsky, C. Esebbag, P. Schuck, Phys. Rev. Lett. 
	{\bf 87} 066403 (2001). 
%
\bibitem{POGHOSSIAN} J.~von Delft and R.~Poghossian, cond-mat/0106405.
%
\bibitem{HIKAMI} K. Hikami, P.P.  Kulish, and  M. Wadati, 
	Jou. Phys. Soc. Jap. {\bf 61},  3071 (1992).
%
\bibitem{BRAUN} F. Braun and J. von Delft, in {\it Quantum mesoscopic phenomena 
	and mesoscopic devices in microelectronics}, NATO, edited by 
	I.O. Kulik and R. Ellialtioglu (Kluwer, Dortech 2000).
%
\bibitem{2D-ELECTRO} M. Lavrentiev and  B. Chabat, 
{\it Methodes de la theorie des fonctions d'une variable complexe}, (MIR, Moscou 1972).
%
\bibitem{RICH-NUM} R.W. Richardson, Phys. Rev. {\bf 141}, 949 (1966).
%
\bibitem{RICCATI} A.G. Ushveridze, Sov. J. Nucl. {\bf 20} 1185 (1989); 
{\it hep-th/9411035}; {\it hep-th/9708059}. 
%
\bibitem{FUTURE} A.~Di~Lorenzo and A.~Mastellone, \emph{in preparation}. 

\end{thebibliography}
\end{document}